\documentclass[letterpaper, 10 pt, conference]{ieeeconf}
\IEEEoverridecommandlockouts

\usepackage{amsmath,amssymb,amsfonts,epsf,epsfig,times}
\usepackage[all]{xy}
\usepackage{graphicx,color}
\usepackage{subfigure}
\usepackage{url}
\usepackage{cite}
\usepackage{tikz}
\usepackage{epstopdf}

\usepackage{keyval,times}

\newtheorem{theorem}{Theorem}[section]
\newtheorem{lemma}{Lemma}[section]
\def\proof{\noindent{\it Proof: }}
\def\QED{\mbox{\rule[0pt]{1.5ex}{1.5ex}}}
\def\endproof{\hspace*{\fill}~\QED\par\endtrivlist\unskip}

\newcommand{\abs}[1]{\left|#1\right|}

\newcommand{\defeq}{\stackrel{\triangle}{=}}

\newtheorem{definition}[theorem]{Definition}

\newtheorem{myremark}[theorem]{Remark}

\newcommand{\Vcal}{\mathcal{V}}

\newcommand{\OMIT}[1]{}

\begin{document}
\title{Circumnavigation of an Unknown Target Using UAVs with\\ Range and Range Rate Measurements}
\author{Yongcan Cao, Jonathan Muse, David Casbeer, and Derek Kingston
\thanks{The authors are with the Control Science Center of Excellence, Air Force Research Laboratory, Wright-Patterson AFB, OH 45433. Approved for public release; distribution unlimited, 88ABW-2013-1085.}
\thanks{Corresponding Author: Yongcan Cao (yongcan.cao@gmail.com)}
}

\markboth{}
         {}

\maketitle

\begin{abstract}
This paper presents two control algorithms enabling a UAV to circumnavigate an unknown target using range and range rate (\textit{i.e.,} the derivative of range) measurements. Given a prescribed orbit radius, both control algorithms (i) tend to drive the UAV toward the tangent of prescribed orbit when the UAV is outside or on the orbit, and (ii) apply zero control input if the UAV is inside the desired orbit.  The algorithms differ in that, the first algorithm is smooth and unsaturated while the second algorithm is non-smooth and saturated.  By analyzing properties associated with the bearing angle of the UAV relative to the target and through proper design of Lyapunov functions, it is shown that both algorithms produce the desired orbit for an arbitrary initial state. Three examples are provided as a proof of concept.
\end{abstract}

\begin{keywords}
Circumnavigation, UAV
\end{keywords}

\IEEEpeerreviewmaketitle

\section{Introduction}

The use of Unmanned Aerial Vehicles (UAVs) in both civilian and military applications has blossomed due to the advancements in aerospace technology. Because they can potentially be built smaller, lighter, and cheaper, UAVs have promising benefits over traditional manned aircraft, yet a key technological challenge remains in designing proper control strategies that will provide a specific degree of autonomy~\cite{Dahm10}. The promise of autonomy has benefits of reduced costs in human management and maintaining a robust and stable performance.

One typical application of UAVs is the surveillance and reconnaissance mission~\cite{TangOzguner05,Kingston07,ShamesDFA12,DeghatSAY13}. The objective is to gather and manage information from various sensors. In a broad sense, a successful surveillance and reconnaissance mission is able to improve situation awareness in an unknown environment through information acquisition. For instance, from the information perspective, a UAV can be deployed to obtain valuable information regarding a target if it can orbit around this target at a desired distance. This type of circular motion around a target is called \emph{circumnavigation}~\cite{ShamesDFA12,DeghatSAY13}. In~\cite{ShamesDFA12}, the circumnavigation problem using range measurements was solved under a unified localization-and-control framework, where two algorithms, namely localization and control, were proposed to guarantee stability. In particular, the concept of \textit{persistent excitation (p.e.)} plays a crucial role in the stability analysis. When the p.e. condition is satisfied, exponential convergence of the localization algorithm and the control algorithm is always guaranteed. In~\cite{DeghatSAY13}, the circumnavigation problem was studied by assuming the availability of bearing measurements. A similar localization-and-control framework was used with the aid of the p.e. concept. In addition to the different measurement types, another major difference between~\cite{ShamesDFA12} and~\cite{DeghatSAY13} is the dynamics used to model the agent. In particular, the single-integrator kinematics was used in~\cite{ShamesDFA12} while the unicycle model was used in~\cite{DeghatSAY13}. One common feature between~\cite{ShamesDFA12} and~\cite{DeghatSAY13} is that the (accurate) location of the UAV is needed to circumnavigate the target.

In an ideal situation, various measurements, such as bearing and location, can be used in the controller design for the circumnavigation mission. However, in some adversarial situations, such as GPS-denied environments due to jamming~\cite{Warwick11} or spoofing~\cite{ShepardBhattiHumphreys12}, measurements become more limited. For instance, under GPS-denied environments, range measurements are possible~\cite{SahinogluGezici06} while other measurements such as bearing and location are not possible. It is thus challenging to design proper controller algorithms with measurement limitations. This is the main motivation of the paper. Upon solving the problem successfully, our subsequent objective is to consider more general navigation and control scenario under GPS-denied environments. 

This paper considers the circumnavigation problem when (1) the unicycle model is used to model the dynamics of UAVs and (2) only range and range rate measurements are available. The consideration of the unicycle model is more appropriate considering the dynamics of UAVs, at the cost of complex control algorithm design and stability analysis due to the nature of nonlinearity and the underactuated system dynamics. The consideration of range and range rate measurements are meaningful when the UAVs have limited information regarding the target and itself due to adversarial environments, such as GPS-denied environments. Some related work was reported in~\cite{MatveevTeimooriSavkin09,ChaudharySinha12} under these two assumptions. In particular, the target following problem, whose objective is to have a robot orbit around a moving target with some desired radius, was considered in~\cite{MatveevTeimooriSavkin09}. The proposed sliding mode controller can guarantee stability when the robot is out of some non-empty domain. In other words, no global stability is guaranteed. In~\cite{ChaudharySinha12}, the problem of detecting a target was considered where the objective is to drive the robot close to a stationary target. Since no special motion of the robot is needed upon approaching the target, the detecting problem considered in~\cite{ChaudharySinha12} is less challenging than the circumnavigation problem. The novel contributions of the paper include a complete global stability analysis for circumnavigation of an unknown target using only range and range rate measurements. Related work includes algorithms for reaching targets with using only range and range rate measurements (but no circumnavigation), or circumnavigation using bearing measurements, or circumnavigation without a guarantee of global stability. To our best knowledge, achieving circumnavigation globally using only range and range rate measurements has not been demonstrated previously. 

The paper is organized as follows. In Section~\ref{sec:problem}, the problem studied in this paper is introduced. A control algorithm is then proposed in Section~\ref{sec:algorithm}. Sections~\ref{sec:stability} and~\ref{sec:proof} analyze, respectively, the stable circular motion using the proposed control algorithm and the conditions under which the stable circular motion can be achieved. In Section~\ref{sec:newAlg}, another control algorithm is proposed and analyzed. Both the algorithm presented in Section~\ref{sec:algorithm} and the algorithm presented in~\ref{sec:newAlg} have unique features. In Section~\ref{sec:simulation}, three simulation examples are provided as a demonstration of the effectiveness of the proposed control algorithms. Section~\ref{sec:conclusion} is a brief summarization of the paper with a short discussion of future research directions.

\section{Problem Statement}
\label{sec:problem}

In this paper, the dynamics of the UAV are modeled as
\begin{align}\label{eq:dynamics}
\dot{x}&=V\cos(\psi),\notag\\
\dot{y}&=V\sin(\psi),\quad t\geq 0\\
\dot{\psi}&=\omega,\notag
\end{align}
where $[x,y]^T$ is the location of the UAV, $V$ is the velocity of the UAV, and $\psi$ is the heading of the UAV. Let there be a static unknown target $T$. The objective is to design $\omega$ such that the UAV can circumnavigate the unknown target at some desired distance $r_d$ using range and range rate measurements. Two assumptions are made regarding the UAV and its measurement capabilities:
\begin{itemize}
\item[1.] The velocity $V$ is constant; and
\item[2.] Both range and range rate are measurable.
\end{itemize}

In~\cite{DeghatSAY13}, the circumnavigation problem was solved by assuming the availability of the bearing angle and the location of the UAV. By estimating the position of the target, the control algorithm was built based on the estimated position of the target. In Section~\ref{sec:algorithm}, a novel control algorithm is proposed based on range and range rate measurements without estimating the position of the target.

\section{Algorithm}\label{sec:algorithm}
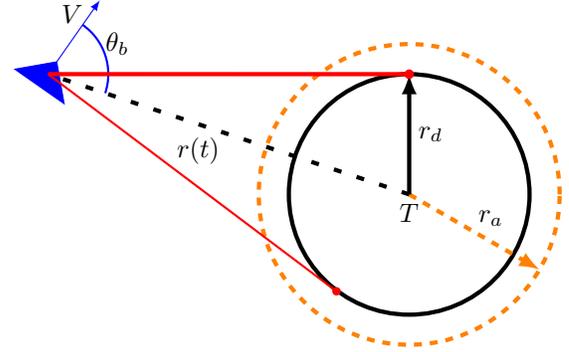
\begin{figure}
\centering
\begin{tikzpicture}[scale=.8]

\coordinate (center) at (8,3);
\coordinate (uav) at (2,5);
\draw [orange, dashed, ultra thick] (center) circle [radius = 2.5];
\draw [black, ultra thick] (center) circle [radius = 2.0];

\draw [-latex, orange, dashed, ultra thick] (center) -- +(-30 : 2.5);
\draw[-latex, black, ultra thick] (center) -- +(90:2);
\node [right] at (9,2.6) {$r_{a}$};
\node [right] at (8,4) {$r_{d}$};
\node [below] at (center) {$T$};

\draw [blue, fill=blue, rotate=-35, shift=(uav)] ++(90:.25) -- ++(-45:.707) -- ++(180:1) -- ++(45:.707);
\draw [black, loosely dashed, ultra thick] (center) -- (uav);
\node [below] at (4.5,4.1) {$r(t)$};
\draw [-latex, blue] (uav) -- ++(55:1.5);
\draw [blue,thick] (uav) +(55:1) arc [radius=1, start angle=55, end angle=-18.435];
\node [] at (3.15,5.5) {$\theta_{b}$};
\node [] at (2.4,6) {$V$};

\draw [red, ultra thick] (uav) -- (8,5);
\draw [red, fill=red] (8,5) circle [radius=2pt];
\draw [red, thick] (uav) -- +(-37:6) [red, fill=red] circle [radius=1.5pt];


\end{tikzpicture}
\caption{An illustration of variables and notations used in the paper. The blue triangle denotes the UAV. $T$ denotes the target. The blue arrow denotes the heading of the UAV. $r(t)$ denotes the range between the UAV and the target. $V$ denotes the (constant) velocity of the UAV. $\theta_b$ denotes the bearing angle. $r_d$ and $r_a$ denotes, respectively, the desired radius and the actual radius.}
\label{fig:motivation}
\end{figure}

Before describing the proposed control algorithm, let's first take a look at a typical scenario in Fig.~\ref{fig:motivation} where the UAV is outside the black solid circle with a radius $r_d$ centering at the target $T$. The proposed control algorithm is a feedback control law by comparing the difference between the desired change rate of $r^2(t)$ and the actual change rate of $r^2(t)$. The desired change rate of $r^2(t)$ is the change rate of $r^2(t)$ when the UAV moves towards the tangent point on the black solid circle. By computation, the desired change rate of $r^2(t)$ is given by $2r(t)V\cos(\pi-\sin^{-1}(\frac{r_d}{r(t)}))$. The actual change rate of $r^2(t)$ is the change rate of $r^2(t)$ when the UAV moves along its current heading. Note that the actual change rate of $r^2(t)$ is given by $2r(t)\dot{r}(t)$. Therefore, when the UAV is outside the black solid circle, $\omega=k[2r(t)V\cos(\pi-\sin^{-1}(\frac{r_d}{r(t)}))-2r(t)\dot{r}(t)]$, where $k$ is a positive constant. When the UAV is inside the black solid circle, no control action is applied to the UAV. The main purpose is to drive the UAV outside the black solid circle. As a summarization, the proposed control algorithm for $\omega$ is given by
\begin{equation}\label{eq:omega}
\left\{
\begin{array} {ll}
k[2r(t)V\cos(\pi-\sin^{-1}(\frac{r_d}{r(t)}))-2r(t)\dot{r}(t)],&r(t)\geq r_d,\\
0,&\text{otherwise},
\end{array}\right.
\end{equation}
where $k$ is a positive constant.\footnote{In practical implementation, $\dot{r}(t)$ can be estimated effectively using a stable linear filter.} Based on the control algorithm~\eqref{eq:omega}, when $r(t)<r_d$, the UAV will keep its current course until the condition $r(t)\geq r_d$ is triggered. When $r(t)< r_d$, the condition $r(t)\geq r_d$ is always triggered after a finite period of time since the UAV is moving along a line with a nonzero velocity. As a consequence, the key is to study the case when $r(t)\geq r_d$, which is the focus of the stability analysis in Section~\ref{sec:proof}.

\section{Stable Motion}\label{sec:stability}
Since the focus is on the circular motion of the UAV around some unknown target, it is assumed that such a stable circular motion exists. Here the stable circular motion is defined as follows.

\begin{definition}
A stable circular motion refers to the behavior that the UAV, with dynamics~\eqref{eq:dynamics}, moves around a target with a constant speed and a constant radius.
\end{definition}

In order to characterize a circular motion, three elements, namely, the center, the direction of rotation, and the radius are needed. The location of the target is the center of the orbit. In the following of the section, the focus is on deriving the direction of rotation and the radius.

Let's first derive the radius. Note that the radius cannot be smaller than $r_d$ due to the control algorithm for the case $r(t)<r_d$. Let the radius of the stable circular motion be given by $r_a(\geq r_d)$. Then the magnitude of the nominal angular velocity is given by
\begin{equation}\label{eq:Vr_a}
\omega^\star = \frac{V}{r_a}.
\end{equation}
In other words, the direction of rotation is either clockwise or counterclockwise. Since $r_a\geq r_d$, it follows from~\eqref{eq:omega} that
\begin{equation}\label{eq:equilibrium-eqt}
\omega=2kr_a V\cos(\pi-\sin^{-1}(\frac{r_d}{r_a})),
\end{equation}
where $\dot{r}_d=0$ since the radius is equal to $r_a$, which is a positive constant. Under the assumption that a stable circular motion exists, it follows that the magnitude of $\omega$ should match $\omega^\star$. Equivalently, it can be obtained that
\begin{equation}\label{eq:omega_krv}
\omega^\star=\abs{\omega}=2kr_a V\abs{\cos(\pi-\sin^{-1}(\frac{r_d}{r_a}))}.
\end{equation}
By definition,
\begin{align}\label{eq:cos-out}
&\abs{\cos(\pi-\sin^{-1}(\frac{r_d}{r_a}))}\notag\\
=&\cos(\sin^{-1}(\frac{r_d}{r_a}))\notag\\
=&\sqrt{1-\frac{r^2_d}{(r_a)^2}}.
\end{align}
By substituting~\eqref{eq:cos-out} into~\eqref{eq:omega_krv}, one can obtain
\begin{equation}
\frac{V}{r_a}=2kr_a V\sqrt{1-\frac{r^2_d}{(r_a)^2}},
\end{equation}
which implies that
\begin{equation}
\frac{1}{4k^2}=(r_a)^4\left(1-\frac{r^2_d}{(r_a)^2}\right).
\end{equation}
By computation, $r_a$ is given by
\begin{equation}\label{eq:r_a}
r_a = \sqrt{\frac{r^2_d+\sqrt{r^4_d+1/k^2}}{2}}.
\end{equation}
Therefore, if a stable circular motion exists, the radius is given by $r_a$ defined in~\eqref{eq:r_a}.

Next, the direction of rotation is derived. Note that the magnitude of $\omega$, \textit{i.e.,} $\frac{V}{r_a}$, denotes the nominal angular velocity and the sign of $\omega$ when $r(t)\equiv r_a$ denotes the direction of rotation. More specifically, if the sign of $\omega$ is positive, the heading angle increases, indicating that the UAV rotates counter clockwise. Analogously, if the sign of $\omega$ is negative, the UAV rotates clockwise. From~\eqref{eq:equilibrium-eqt}, the sign of $\omega$ is determined by the sign of $\cos(\pi-\sin^{-1}(\frac{r_d}{r_a}))$. Since $r_a> r_d$ and $\sin^{-1}(\frac{r_d}{r_a})\in(0,\frac{\pi}{2})$, $\pi-\sin^{-1}(\frac{r_d}{r_a})\in(\frac{\pi}{2},\pi)$. It follows that $\cos(\pi-\sin^{-1}(\frac{r_d}{r_a}))<0$. Combining with the previous analysis implies that the UAV rotates clockwise when $k>0$. 

As a summarization, the following theorem illustrates the properties of the stable circular motion.
\begin{theorem}\label{th:circular-property}
Consider system dynamics~\eqref{eq:dynamics} subject to control input~\eqref{eq:omega}. If a stable circular motion exists, the radius is given by $r_a$ in~\eqref{eq:r_a}. In addition, the UAV rotates clockwise when $k>0$.
\end{theorem}

From the previous analysis, it can be observed that the actual radius $r_a$ is greater than the desired radius $r_d$. In order to guarantee that $r_a$ matches the desired distance, one can choose $r_d$ as
\begin{equation}\label{eq:tilde_r_d}
\tilde{r}_d\defeq r_d\sqrt{1-\frac{1}{4k^2r_d^4}}.
\end{equation}
By using $\tilde{r}_d$ to replace $r_d$ in~\eqref{eq:omega}, it can be computed from~\eqref{eq:r_a} that $r_a=r_d$, which indeed meets our objective as mentioned in Section~\ref{sec:problem}. To guarantee the existence of a positive $\tilde{r}_d$, it is required that $k>\frac{1}{2r_d^2}$.

Up to now, the property of the stable circular motion using~\eqref{eq:omega} was analyzed under the assumption that the stable circular motion does exist. However, it remains unclear if the assumption stands. In Section~\ref{sec:proof}, it is shown that this assumption always holds for any nonzero $k$ satisfying $k>\frac{1}{2r_d^2}$.

\section{Main Result}\label{sec:proof}

This section proves the main result of the paper given in the following theorem:

\begin{theorem}\label{th:circular}
Consider the UAV dynamics in~\eqref{eq:dynamics} subject to the control policy in~\eqref{eq:omega}. If $k>\frac{1}{2r_d^2}$, then $r(t) \to r_a$ as $t \to \infty$, where $r_a$ is defined in~\eqref{eq:r_a}.
\end{theorem}

Before proving Theorem~\ref{th:circular}, a few definitions and lemmas are necessary.
\begin{definition}\label{de:C_d}
The circle centered at the unknown target with a radius $r_d$ is defined as $C_d$. The UAV is inside (respectively, outside) $C_d$ if $r(t)<r_d$ (respectively, $r(t)\geq r_d$).
\end{definition}
\begin{definition}\label{de:bearing}
Denote the reference vector as the vector from the current location of the UAV to the target. The bearing angle $\theta_b(t)\in[0,2\pi)$ at time $t$ is defined as the angle from the reference vector to the current heading of the UAV measured counterclockwise.
\end{definition}

As an example, in Fig.~\ref{fig:motivation}, the black solid circle denotes $C_d$ as defined in Definition~\ref{de:C_d}. An illustration of the bearing angle $\theta_b$ is also shown in this figure.

\begin{lemma}\label{lem:not_in_C_d}
Consider the UAV dynamics in~\eqref{eq:dynamics} subject to the control policy in~\eqref{eq:omega}. Let there be $t_0\geq 0$ such that $r(t_0) \geq r_d$ and $\theta_b(t_0)\in(\sin^{-1}(\frac{r_d}{r(t_0)}),2\pi-\sin^{-1}(\frac{r_d}{r(t_0)}))$, then $r(t) \geq r_d,$ $\forall t \geq t_0$.
\end{lemma}
\proof 
The proof of the lemma can be divided into the following two steps:

Step 1: $\theta_b(t)\in[\sin^{-1}(\frac{r_d}{r(t)}),2\pi-\sin^{-1}(\frac{r_d}{r(t)}))$ holds for all $t\geq t_0$. Based on the proposed algorithm~\eqref{eq:omega}, $\omega<0$ at $t=t_0$ since $2r(t)\dot{r}(t)>2r(t)V\cos(\pi-\sin^{-1}(\frac{r_d}{r(t)}))$ when $t=t_0$. As a consequence, the UAV will rotate clockwise initially. Note that both $2r(t)\dot{r}(t)$ and $2r(t)V\cos(\pi-\sin^{-1}(\frac{r_d}{r(t)}))$ are continuous with respect to $t$. Therefore, $2r(t)\dot{r}(t)>2r(t)V\cos(\pi-\sin^{-1}(\frac{r_d}{r(t)}))$ always holds before $2r(t)\dot{r}(t)=2r(t)V\cos(\pi-\sin^{-1}(\frac{r_d}{r(t)}))$ happens. Consequently, the UAV will stop rotating clockwise once $2r(t)\dot{r}(t)=2r(t)V\cos(\pi-\sin^{-1}(\frac{r_d}{r(t)}))$. Indeed, $2r(t)\dot{r}(t)=2r(t)V\cos(\pi-\sin^{-1}(\frac{r_d}{r(t)}))$ if and only if $\theta_b(t)=\sin^{-1}(\frac{r_d}{r(t)})$ or $\theta_b(t)=2\pi-\sin^{-1}(\frac{r_d}{r(t)})$. Thus, $\theta_b(t)\in[\sin^{-1}(\frac{r_d}{r(t)}),2\pi-\sin^{-1}(\frac{r_d}{r(t)})]$. To prove Step 1, it suffices to show that $\theta_b(t)=2\pi-\sin^{-1}(\frac{r_d}{r(t)})$ cannot hold. When $\theta_b(t_0)\in(\sin^{-1}(\frac{r_d}{r(t_0)}),2\pi-\sin^{-1}(\frac{r_d}{r(t_0)}))$ and $\omega= 0$, $\theta_b(t)=2\pi-\sin^{-1}(\frac{r_d}{r(t)})$ never holds for all $t\geq t_0$ because the UAV moves along a straight line and the straight line is always outside the circle $C_d$. As a consequence, $\theta_b(t)=2\pi-\sin^{-1}(\frac{r_d}{r(t)})$ never holds for all $t\geq t_0$ when $\omega\leq 0$ and $\theta_b(t_0)\in[\sin^{-1}(\frac{r_d}{r(t_0)}),2\pi-\sin^{-1}(\frac{r_d}{r(t_0)}))$. Combining the previous arguments completes the proof of Step 1.

Step 2: $r(t)\geq r_d$ for all $t\geq t_0$. From Step 1, it is known that $\theta_b(t)\in[\sin^{-1}(\frac{r_d}{r(t)}),2\pi-\sin^{-1}(\frac{r_d}{r(t)}))$. When $r(t)=r_d$, it follows that $\sin^{-1}(\frac{r_d}{r(t)})=\frac{\pi}{2}$. Combining with the previous two sentences indicates that $\theta_b(t)\in(\frac{\pi}{2},\frac{3\pi}{2}]$ at the time when $r(t)=r_d$. Noticing that
\begin{align}\label{eq:rdot}
\dot{r}=-V\cos(\theta_b(t)),
\end{align}
it then follows that $\dot{r}\geq 0$ at the time when $r(t)=r_d$. This implies that the UAV cannot get any closer to the target if $r(t)=r_d$. At the time $t^\star$ when $r(t^\star)$ becomes larger than $r_d$, the bearing angle has to be in the set $(\frac{\pi}{2},\frac{3\pi}{2})$, which satisfies the condition that $\theta_b(t^\star)\in[\sin^{-1}(\frac{r_d}{r(t^\star)}),2\pi-\sin^{-1}(\frac{r_d}{r(t^\star)}))$. By repeating the analysis in Steps 1 and 2, it is clear that the UAV will never move inside $C_d$.
\endproof

\begin{lemma}\label{lem:in_C_d_once}
Consider the UAV dynamics in~\eqref{eq:dynamics} subject to the control policy in~\eqref{eq:omega}. The UAV can only move inside $C_d$ at most once.
\end{lemma}
\proof When $r(t_e)=r_d$, $\theta_b(t_e)\in[0,\frac{\pi}{2})\cup(\frac{3\pi}{2},2\pi)$ in order to guarantee that the UAV enters $C_d$. Recall that no control input is imposed on the UAV when it is inside $C_d$. Then at the time $t_x$ when it exits $C_d$,
\begin{align*}
\theta_b(t_x)=
\left\{\begin{array} {ll}
\pi-\theta_b(t_e),&\theta_b(t_e)\in[0,\frac{\pi}{2}),\\
3\pi-\theta_b(t_e),&\theta_b(t_e)\in(\frac{3\pi}{2},2\pi),
\end{array}\right.
\end{align*}
because the UAV moves along a straight line due to the fact $\omega=0$. As a consequence, $\theta_b(t_x)\in(\frac{\pi}{2},\frac{3\pi}{2})$. When $r(t_x)=r_d$, it follows that $\sin^{-1}(\frac{r_d}{r(t_x)})=\frac{\pi}{2}$. Therefore, $\theta_b(t_x)\in(\sin^{-1}(\frac{r_d}{r(t_x)}),2\pi-\sin^{-1}(\frac{r_d}{r(t_x)}))$ when $r(t_x)=r_d$. By considering the current time $t_x$ be the time $t_0$ in Lemma~\ref{lem:not_in_C_d}, it follows from Lemma~\ref{lem:not_in_C_d} that the UAV will never move inside $C_d$ again. Therefore, the UAV can only move inside $C_d$ at most once.
\endproof

\begin{lemma}\label{lem:zero_theta_pi}
Consider the UAV dynamics in~\eqref{eq:dynamics} subject to the control policy in~\eqref{eq:omega}. For any $\theta_b(0)$, there exists $t^\star\geq 0$ such that $\theta_b(t)\in[0,\pi]$ for any $t\geq t^\star$.
\end{lemma}
\proof By Lemma~\ref{lem:in_C_d_once}, the UAV can move inside $C_d$ at most once. When the UAV never moves inside $C_d$, let $t_1=0$. When the UAV moves inside $C_d$ once, let $t_1$ be the time when the UAV moves from inside $C_d$ to outside $C_d$. It is clear that $t_1$ is finite. The lemma is proved if the following two statements are valid:\newline
(1) For any $\theta_b(0)$, there exists $t^\star\geq t_1$ such that $\theta_b(t^\star)\in[0,\pi]$; and\newline
(2) Once $\theta_b(t^\star)\in[0,\pi]$ for some $t^\star\geq t_1$, $\theta_b(t)\in[0,\pi]$ for any $t\geq t^\star$.

The first statement is proved by considering the following three cases:
\begin{itemize}
\item[(i)] $\theta_b(t_1)\in(\pi,2\pi-\sin^{-1}(\frac{r_d}{r(t_1)}))$: As defined in Definition~\ref{de:bearing}, it can be obtained that
\begin{align}\label{eq:thetadot}
\dot{\theta}_b(t)=\omega-\left(-\frac{V\sin(\theta_b(t))}{r(t)}\right),
\end{align}
where $-\frac{V\sin(\theta_b(t))}{r(t)}$ is the angular velocity of the reference vector and $\omega$ denotes the angular velocity of the heading.
Therefore, $\dot{\theta}_b(t)<0$, \textit{i.e.,} $\theta_b(t)$ will decrease, whenever $\theta_b(t)\in(\pi,2\pi-\sin^{-1}(\frac{r_d}{r(t)}))$ because $\omega<0$ while $-\frac{V\sin(\theta_b(t))}{r(t)}>0$. When $\omega$ is (approximately) zero, $-\frac{V\sin(\theta_b(t))}{r(t)}$ is (approximately) $\frac{Vr_d}{r^2(t)}$. Because both $\omega$ and $-\frac{V\sin(\theta_b(t))}{r(t)}$ are continuous with respect to $t$, $\dot{\theta}_b(t)$ is always upper bounded by some negative constant. Similarly, when $-\frac{V\sin(\theta_b(t))}{r(t)}$ is (approximately) zero, $\omega$ is upper bounded by some negative constant, indicating that $\dot{\theta}_b(t)$ is always upper bounded by some negative constant as well. Therefore, $\theta_b(t)\leq \pi$ in finite time. Noting that $\theta_b(t_1)\in(\pi,2\pi-\sin^{-1}(\frac{r_d}{r(t_1)}))\in(\sin^{-1}(\frac{r_d}{r(t_1)}),2\pi-\sin^{-1}(\frac{r_d}{r(t_1)}))$, it follows from Step 1 in the proof of Lemma~\ref{lem:not_in_C_d} that $\theta_b(t)$ cannot get smaller than $\sin^{-1}(\frac{r_d}{r(t)})$, which implies that $\theta_b(t)\geq \sin^{-1}(\frac{r_d}{r(t)})$.
\item[(ii)] $\theta_b(t_1)=2\pi-\sin^{-1}(\frac{r_d}{r(t_1)})$: Under this case, the UAV is heading towards the tangent point such that the bearing is $2\pi-\sin^{-1}(\frac{r_d}{r(t_1)})$. By computation, $\omega=0$, which implies that the UAV will move along a straight line towards the tangent point. As a consequence, the UAV will move towards the tangent point whenever $r(t)>r_d$. Given a constant nonzero velocity, it takes a finite period of time before $r(t)=r_d$ happens. Notice that $r(t)=r_d$ cannot hold for an arbitrary period of time because otherwise a contradiction happens by noting that (i) $\omega=0$ based on~\eqref{eq:omega} during that period of time, indicating that the UAV cannot rotate; and (ii) the UAV has to rotate such that $r(t)=r_d$ holds for that period of time. This implies that $r(t)$ will increase to be greater than $r_d$ as soon as $r(t)=r_d$ happens. The bearing angle $\theta_b(t_f)$ must be in the interval $(\frac{\pi}{2},\frac{3\pi}{2})$ at the time when $r(t)$ increases to be greater than $r_d$. When $\theta_b(t_f) \in(\pi,\frac{3\pi}{2})$, it follows from Case (i) that $\theta_b(t)$ will be in the set $[0,\pi]$ after a finite period of time. When $\theta_b(t_f) \in(\frac{\pi}{2},\pi]$, it is already in the set $[0,\pi]$.
\item[(iii)] $\theta_b(t_1)=(2\pi-\sin^{-1}(\frac{r_d}{r(t_1)}),2\pi)$: If $\theta_b(t)=(2\pi-\sin^{-1}(\frac{r_d}{r(t)}),2\pi)$ always holds, it takes a finite period of time before $r(t)\leq r_d$ happens. Since the UAV never moves inside $C_d$ for $t\geq t_1$, either Case (i) or Case (ii) will happen after a finite period of time. By following the analysis in Cases (i) and (ii), $\theta_b(t)$ will be in the set $[0,\pi]$ after a finite period of time.
\end{itemize}
To prove the second statement, it is essential to study $\dot{\theta}_b(t)$ when $\theta_b(t)=0$ or $\theta_b(t)=\pi$.
Whenever $\theta_b(t)=0$ or $\theta_b(t)=\pi$, the change rate of the reference vector defined in Definition~\ref{de:bearing} is zero. Then it follows that $\dot{\theta}_b(t)=\omega$. When $\theta_b(t)=0$, it can be computed that \[\omega=k[2r(t)V\cos(\pi-\sin^{-1}(\frac{r_d}{r(t)}))+2r(t)V]>0,\]
where~\eqref{eq:rdot} was used to derive the equality. This indicates that $\theta_b(t)$ will increase as soon as $\theta_b(t)=0$ happens. Similarly, when $\theta_b(t)=\pi$, one can obtain that
\[\omega=k[2r(t)V\cos(\pi-\sin^{-1}(\frac{r_d}{r(t)}))-2r(t)V]<0,\]
which indicates that $\theta_b(t)$ will decrease as soon as $\theta_b(t)=\pi$ happens.  Therefore the second statement holds as well.
\endproof

With the previous lemmas, we next prove Theorem~\ref{th:circular} restated as:\newline
\textit{Theorem~\ref{th:circular}:}
Consider the UAV dynamics in~\eqref{eq:dynamics} subject to the control policy in~\eqref{eq:omega}. If $k>\frac{1}{2r_d^2}$, then $r(t) \to r_a$ as $t \to \infty$, where $r_a$ is defined in~\eqref{eq:r_a}.

\proof
Based on Lemmas~\ref{lem:in_C_d_once} and~\ref{lem:zero_theta_pi}, there exists a time instant $t^\star$ such that $\theta_b(t)\in[0,\pi]$ and $r(t)\geq r_d$ for any $t\geq t^\star$. Therefore, $\theta_b(t)$ is continuously differentiable with respect to $t$ for $t\geq t^\star$. In addition, $r(t^\star)$ remains bounded because the velocity of the UAV is bounded. For $t\geq t^\star$, consider the following Lyapunov function candidate given by
\begin{align*}
\Vcal = 1-\sin(\theta_b(t))+\varphi,
\end{align*}
where
$
\varphi=
\int_{r_a}^r (\frac{1}{r_a}-\frac{1}{z}+2kz\cos\sin^{-1}(\frac{r_d}{z})-2kr_a\cos\sin^{-1}(\frac{r_d}{r_a}))\text{d}z\geq 0.
$
Note that $\Vcal$ is continuous with respect to $t$ and $\Vcal\geq 0$. From~\eqref{eq:Vr_a} and~\eqref{eq:omega_krv}, one can obtain that $\frac{1}{r_a}=2kr_a\cos\sin^{-1}(\frac{r_d}{r_a})$. Then $\Vcal$ can be rewritten as
\begin{align*}
\Vcal = 1-\sin(\theta_b(t))+\int_{r_a}^r (-\frac{1}{z}+2kz\cos\sin^{-1}(\frac{r_d}{z}))\text{d}z.
\end{align*}
For $t\geq t^\star$, the derivative of $\Vcal$ along the solution of~\eqref{eq:dynamics} using~\eqref{eq:omega} is given by
\begin{align*}
\dot{\Vcal}&=-\cos(\theta_b(t))\dot{\theta}_b(t)+\dot{\varphi}\\
&=-\cos(\theta_b(t))\Bigg[2krV\left(\cos(\pi-\sin^{-1}(\frac{r_d}{r}))+\cos(\theta_b(t))\right)\\
&~~~~~~~~~~~~~~~~~~~~~+\frac{V\sin(\theta_b(t))}{r}\Bigg]\\
&~~~-\left[2kr\cos\sin^{-1}(\frac{r_d}{r})-\frac{1}{r}\right]V\cos(\theta_b(t))\\
&=V\cos(\theta_b(t))\left[-2kr\cos(\theta_b(t))-\frac{\sin(\theta_b(t))}{r}+\frac{1}{r}\right],
\end{align*}
where~\eqref{eq:rdot} and~\eqref{eq:thetadot} were used to derive the second equality.
When $\theta_b(t)\in[\frac{\pi}{2},\pi]$, $\dot{\Vcal}\leq 0$ because $\cos(\theta_b(t))\leq 0$ and $\sin(\theta_b(t))\leq 1$. When $\theta_b(t)\in[0,\frac{\pi}{2})$, $\dot{\Vcal} < 0$ if $-2kr\cos(\theta_b(t))-\frac{\sin(\theta_b(t))}{r}+\frac{1}{r}<0$. When $k>\frac{1}{2r_d^2}$, it follows from the fact $r\geq r_d$ that $2kr^2>1$. Therefore,
\begin{align*}
&-2kr\cos(\theta_b(t))-\frac{\sin(\theta_b(t))}{r}+\frac{1}{r}\\
<&-\frac{\cos(\theta_b(t))}{r}-\frac{\sin(\theta_b(t))}{r}+\frac{1}{r}\leq 0.
\end{align*}
Therefore, $\dot{\Vcal}\leq 0$ for $\theta_b(t)\in[0,\pi]$. Because $\dot{\Vcal}$ is uniformly continuous when $r\geq r_d$ and $\theta_b(t)\in[0,\pi]$, it follows from Lemma 4.3 in~\cite{SlotineLi91} that $\dot{\Vcal}\to 0$ as $t\to\infty$. When $k>\frac{1}{2r_d^2}$, $\dot{\Vcal}=0$ implies that $\theta_b=\frac{\pi}{2}$. It then follows from~\eqref{eq:rdot} that when $\theta_b(t)=\frac{\pi}{2}$, $r(t)$ is constant. It then follows from the analysis in Section~\ref{sec:stability} that $r(t)=r_a$. Therefore, $\theta_b(t)\to\frac{\pi}{2}$ and $r(t)\to r_a$ as $t\to\infty$.
\endproof

When $\theta_b(t)\in[0,2\pi)$ rather than $\theta_b(t)\in[0,\pi)$, $\dot{\Vcal}$ is not necessarily negative semi-definite. In particular, when $\theta_b(t)$ is just a bit larger than $\frac{3\pi}{2}$, $\dot{\Vcal}$ becomes positive for some bounded $k$ and $r$. Therefore, Lemma~\ref{lem:zero_theta_pi}  is essential in the proof of Theorem~\ref{th:circular}.

Note from Theorem~\ref{th:circular} that $\theta_b(t)\to\frac{\pi}{2}$ and $r(t)\to r_a$ as $t\to\infty$. In order to guarantee that $\theta_b(t)\to\frac{\pi}{2}$ and $r(t)\to r_d$ as $t\to\infty$, one can simply replace $r_d$ by $\tilde{r}_d$ given in~\eqref{eq:tilde_r_d}.

\section{A New Algorithm and Its Stability}\label{sec:newAlg}
Although the algorithm~\eqref{eq:omega} can guarantee a stable circular motion, the radius of the circular motion might not match the desired value. In fact, the difference between them is determined by various parameters. As mentioned in Section~\ref{sec:stability}, one way to guarantee that the desired radius can be reached is to intentionally change $r_d$, which requires a design of $r_d$ \textit{a priori}. We next propose a new algorithm, which does not require the intentional change of $r_d$. That is, $r_d$ is exactly the desired radius.

The proposed control algorithm $\omega$ is given as
\begin{equation}\label{eq:omega-sign-version}
\left\{
\begin{array} {ll}
k\text{sign}(V\cos(\pi-\sin^{-1}(\frac{r_d}{r(t)}))-\dot{r}(t)),&r(t)\geq r_d,\\
0,&\text{otherwise},
\end{array}\right.
\end{equation}
where $\text{sign}(\cdot)$ is the signum function and $k$ is a positive constant.
It can be considered as a non-smooth version of the algorithm~\eqref{eq:omega}.
In addition to the benefit that $r_d$ is chosen as the (exact) desired radius, another (physical) benefit is that the control input is always saturated, meaning that $\abs{\omega}<k$. From the application's perspective, this control algorithm is implementable even if the control algorithm~\eqref{eq:omega} is not implementable in certain cases. In the following part of the section, it is shown that this algorithm can solve the circumnavigation problem as well.

\begin{theorem}\label{th:circular-sign}
Consider the UAV dynamics in~\eqref{eq:dynamics} subject to the control policy in~\eqref{eq:omega-sign-version}. If $k>\frac{V}{r_d}$, then $r(t)\to r_d$ as $t\to\infty$. In particular, the UAV will rotate clockwise around the target ultimately.
\end{theorem}
\proof By following a similar analysis to Lemmas~\ref{lem:not_in_C_d},~\ref{lem:in_C_d_once}, and~\ref{lem:zero_theta_pi}, one can obtain that the results in these lemmas are still valid under the control algorithm~\eqref{eq:omega-sign-version}. By recalling the statements in Lemma~\ref{lem:zero_theta_pi}, for any $\theta_b(0)$, there exists $t^\star\geq 0$ such that $\theta_b(t)\in[0,\pi]$ and $r(t)\geq r_d$ for any $t\geq t^\star$. It is shown next that $\theta_b(t)\in[0,\frac{\pi}{2}]$ in finite time. Similar to the analysis in the proof of Lemma~\ref{lem:zero_theta_pi}, $\theta_b(t)\geq 0$ for all $t\geq t^\star$ because $\dot{\theta}_b(t)>0$ once $\theta_b(t)=0$. Our focus next is to show that $\theta_b(t)\leq \frac{\pi}{2}$ in finite time. For $t\geq t^\star$, since $r(t)\geq r_d$, it follows from~\eqref{eq:thetadot} that
\begin{align*}
\dot{\theta}_b(t)=&\omega+\frac{V\sin(\theta_b(t))}{r(t)}\\
=&k\text{sign}(\cos(\pi-\sin^{-1}(\frac{r_d}{r(t)}))+\cos(\theta_b(t)))\\
&+\frac{V\sin(\theta_b(t))}{r(t)}\\
\leq & k\text{sign}(\cos(\theta_b(t)))+\frac{V}{r_d}
\end{align*}
by noting that $\cos(\pi-\sin^{-1}(\frac{r_d}{r(t)}))\leq 0$ when $r(t)\geq r_d$, $\sin(\theta_b(t))\leq 1$, and $r(t)\geq r_d$. As a consequence, $\dot{\theta}_b(t)<-k+\frac{V}{r_d}$ whenever $\theta_b(t)>\frac{\pi}{2}$. Because $k>\frac{V}{r_d}$, it follows that $\dot{\theta}_b(t)<0$ whenever $\theta_b(t)>\frac{\pi}{2}$. Therefore, it takes a finite period of time before $\theta_b(t)=\frac{\pi}{2}$ happens. In addition, $\theta_b(t)$ is always not greater than $\frac{\pi}{2}$ afterwards.

Because $\theta_b(t)\in[0,\frac{\pi}{2}]$ in finite time, there must exist a positive constant $\overline{t}\geq t^\star$ such that $\theta_b(t)\in[0,\frac{\pi}{2}]$ for all $t\geq \overline{t}$. For $t\geq \overline{t}$, consider the Lyapunov function candidate given by
\begin{equation*}
\Vcal=r(t)-r_d.
\end{equation*}
Because $r(t)\geq r_d$, $\Vcal\geq 0$. In addition, the derivative of $\Vcal$ is given by
\begin{equation*}
\dot{\Vcal}=\dot{r}(t)=-V\cos(\theta_b(t))\leq 0.
\end{equation*}
Note that $\dot{\Vcal}$ is uniformly continuous with respect to $t$. It then follows from Lemma 4.3 in~\cite{SlotineLi91} that $\dot{\Vcal}\to 0$ as $t\to\infty$. That is, $\cos(\theta_b(t))\to 0$ as $t\to\infty$. Equivalently, $\theta_b(t)\to\frac{\pi}{2}$ as $t\to\infty$ by noting that $\theta_b(t)\in[0,\frac{\pi}{2}]$. When $\theta_b(t)=\frac{\pi}{2}$, $r(t)$ remains constant. If the constant is not $r_d$, $\dot{\theta}_b(t)=-k+\frac{V}{r(t)}<0$, which contradicts the fact that $\theta_b(t)=\frac{\pi}{2}$. Therefore, $\theta_b(t)\to\frac{\pi}{2}$ and $r(t)\to r_d$ as $t\to\infty$. This proves the first statement.

The second statement follows from the fact that the bearing angle $\theta_b(t)\in[0,\frac{\pi}{2}]$ in finite time (see the second paragraph of the proof).
\endproof

\begin{myremark}
Until now, it is assumed that $k$ is a positive constant. When $k$ is a negative constant, stable circular motions can be obtained for the UAV dynamics in~\eqref{eq:dynamics} subject to the control algorithms in~\eqref{eq:omega} and~\eqref{eq:omega-sign-version}. In particular, the radius of the stable circular motion remains $r_a$ under the control algorithm~\eqref{eq:omega} if $k<-\frac{1}{2r_d^2}$. The radius of the stable circular motion remains $r_d$ under the control algorithm~\eqref{eq:omega-sign-version} if $k<-\frac{V}{r_d}$. However, the UAV will ultimately rotate counter clockwise around the target.
\end{myremark}

\section{Simulation}\label{sec:simulation}

This section presents three simulation examples. The velocity $V$ is chosen to be $1$. The location of the unknown target is $[0,-10]^T$. The desired radius $r_d$ is chosen as $10$. The initial state of the UAV is randomly chosen from the set $[0,10]\times [0,10]\times[0,2\pi)$. For the control algorithm~\eqref{eq:omega}, $k$ is chosen to be $0.01$. For the control algorithm~\eqref{eq:omega-sign-version}, $k$ is chosen to be $0.12$. It can be verified that the conditions in Theorems~\ref{th:circular} and~\ref{th:circular-sign} are satisfied.

\begin{figure}
\begin{centering}
\includegraphics[width=.5\textwidth]{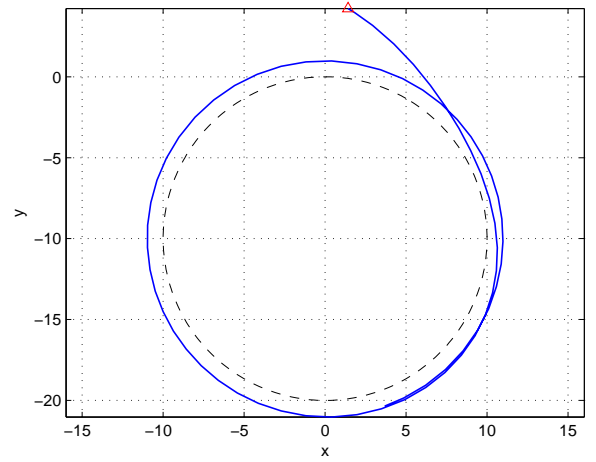}    
\caption{The trajectory of the UAV under~\eqref{eq:omega} with $r_d=10$. The red triangle represents the starting position of the UAV. The dashed line represents the desired trajectory.}  
\label{fig:traj_1}                                 
\end{centering}                                 
\end{figure}

\begin{figure}
\begin{center}
\includegraphics[width=.5\textwidth]{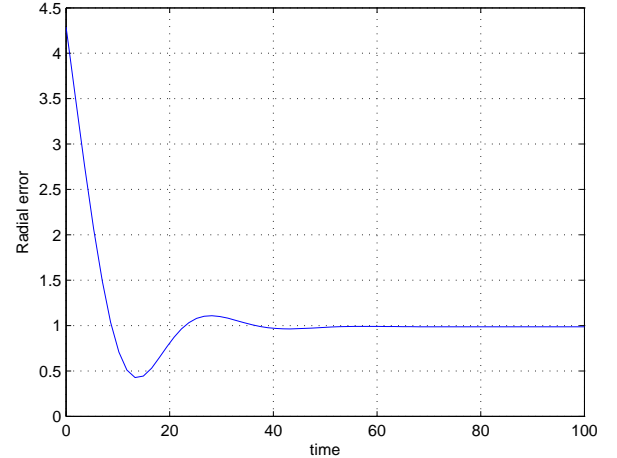}    
\caption{The tracking error of the UAV under~\eqref{eq:omega} with $r_d=10$.}  
\label{fig:error_1}                                 
\end{center}                                 
\end{figure}

Figs.~\ref{fig:traj_1} and~\ref{fig:error_1} show, respectively, the trajectory of the UAV and the tracking error (\textit{i.e.,} $r(t)-r_d$) under the control algorithm~\eqref{eq:omega}. By computation from~\eqref{eq:r_a}, $r_a=10.9868$, which is consistent with the simulation example. It can be observed from Fig.~\ref{fig:traj_1} that a circular motion is obtained ultimately. In addition, the UAV rotates clockwise around the target. Meanwhile, it can be observed that $r(t)\geq r_d$ always holds.

\begin{figure}
\begin{center}
\includegraphics[width=.5\textwidth]{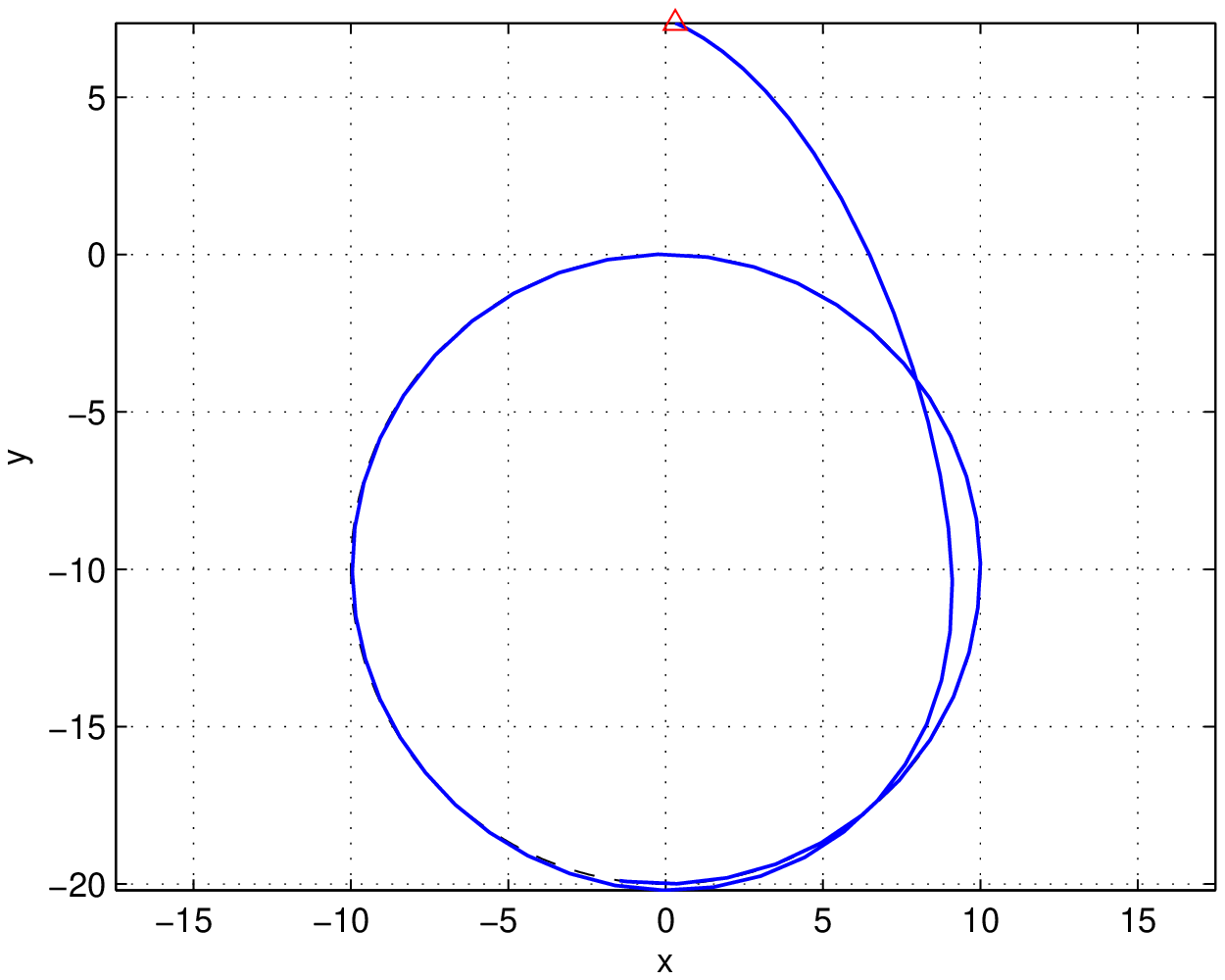}    
\caption{The trajectory of the UAV under~\eqref{eq:omega} with $r_d$ being replaced by $\tilde{r}_d$ in~\eqref{eq:tilde_r_d}. The red triangle represents the starting position of the UAV. The dashed line represents the desired trajectory.}  
\label{fig:traj_1_correction}                                 
\end{center}                                 
\end{figure}

\begin{figure}
\begin{center}
\includegraphics[width=.5\textwidth]{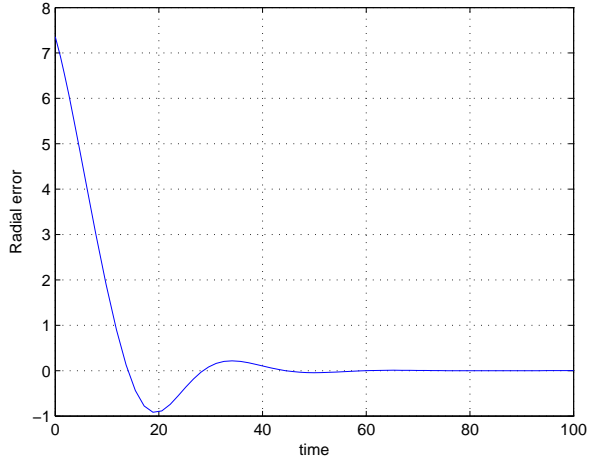}    
\caption{The tracking error of the UAV under~\eqref{eq:omega} with $r_d$ being replaced by $\tilde{r}_d$ in~\eqref{eq:tilde_r_d}.}  
\label{fig:error_1_correction}                                 
\end{center}                                 
\end{figure}

By replacing $r_d$ in~\eqref{eq:omega} by $\tilde{r}_d$ in~\eqref{eq:tilde_r_d}, the trajectory of the UAV and the tracking error under the control algorithm~\eqref{eq:omega} are given in, respectively, Figs.~\ref{fig:traj_1_correction} and~\ref{fig:error_1_correction}. It can be noticed that the actual radius of the circular motion is exactly $10$. Therefore, when $r_d$ is chosen properly, the UAV will circumnavigate around the target at the desired distance.

\begin{figure}
\begin{center}
\includegraphics[width=.5\textwidth]{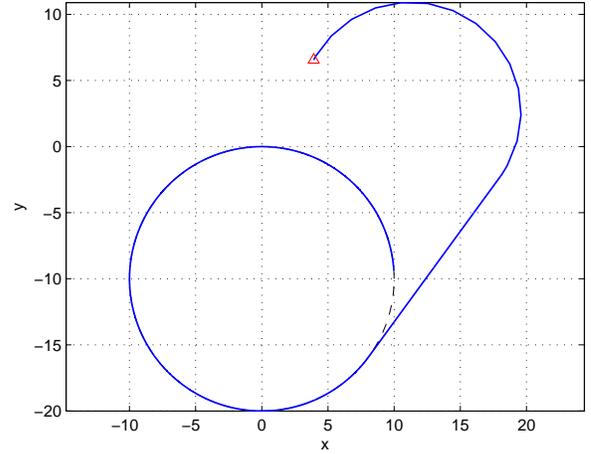}    
\caption{The trajectory of the UAV under~\eqref{eq:omega-sign-version}. The red triangle represents the starting position of the UAV. The dashed line represents the desired trajectory.}  
\label{fig:traj_2}                                 
\end{center}                                 
\end{figure}

\begin{figure}
\begin{center}
\includegraphics[width=.5\textwidth]{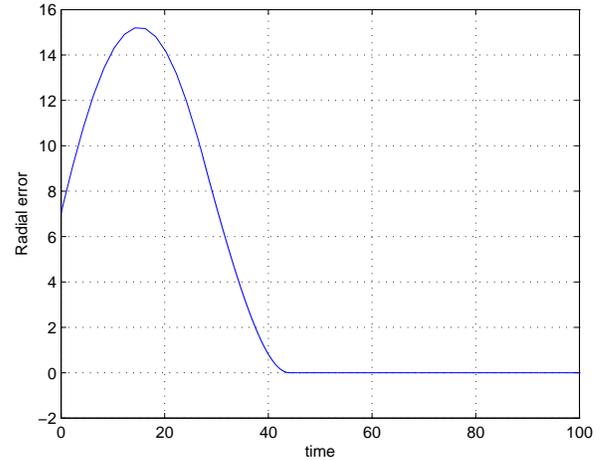}    
\caption{The tracking error of the UAV under~\eqref{eq:omega-sign-version}.}  
\label{fig:error_2}                                 
\end{center}                                 
\end{figure}

Figs.~\ref{fig:traj_2} and~\ref{fig:error_2} show, respectively, the trajectory of the UAV and the tracking error under the control algorithm~\eqref{eq:omega-sign-version}. Again, a circular motion is obtained ultimately. In addition, the final tracking error goes to zero. By comparing the simulation results using~\eqref{eq:omega} and those using~\eqref{eq:omega-sign-version}, it can be seen that the convergence speed using~\eqref{eq:omega} is faster than that using~\eqref{eq:omega-sign-version}. This is mainly due to the fact that the control algorithm~\eqref{eq:omega} is not saturated while the control algorithm~\eqref{eq:omega-sign-version} is saturated.

\section{Conclusion and Future Work}\label{sec:conclusion}

In this paper, we considered the circumnavigation of UAVs around some unknown target at a desired distance using range and range rate measurements. Two control algorithms were proposed to guarantee the accomplishment of the circumnavigation mission. The stability analysis was based on analyzing the properties associated with the bearing angle and designing proper Lyapunov functions. Considering the fact that range and range rate measurements are noisy in practical applications, one future research direction is on analyzing the effect of measurements noises on the performance of the proposed control algorithms. Other future research directions include the study of the circumnavigation mission in the presence of wind, and the study of cooperative circumnavigation mission when multiple UAVs circumnavigate around a target in a cooperative fashion.

\bibliographystyle{ieeetran}
\bibliography{refs}

\begin{thebibliography}{10}
\providecommand{\url}[1]{#1}
\csname url@samestyle\endcsname
\providecommand{\newblock}{\relax}
\providecommand{\bibinfo}[2]{#2}
\providecommand{\BIBentrySTDinterwordspacing}{\spaceskip=0pt\relax}
\providecommand{\BIBentryALTinterwordstretchfactor}{4}
\providecommand{\BIBentryALTinterwordspacing}{\spaceskip=\fontdimen2\font plus
\BIBentryALTinterwordstretchfactor\fontdimen3\font minus
  \fontdimen4\font\relax}
\providecommand{\BIBforeignlanguage}[2]{{%
\expandafter\ifx\csname l@#1\endcsname\relax
\typeout{** WARNING: IEEEtran.bst: No hyphenation pattern has been}%
\typeout{** loaded for the language `#1'. Using the pattern for}%
\typeout{** the default language instead.}%
\else
\language=\csname l@#1\endcsname
\fi
#2}}
\providecommand{\BIBdecl}{\relax}
\BIBdecl

\bibitem{Dahm10}
W.~J. Dahm, ``Technology horizons: A vision for air force science \& technology
  during 2010-2030,'' USAF, Tech. Rep., 2010.

\bibitem{TangOzguner05}
Z.~Tang and U.~Ozguner, ``Motion planning for multitarget surveillance with
  mobile sensor agents,'' \emph{{IEEE} Transactions on Robotics}, vol.~21,
  no.~5, pp. 898--908, 2005.

\bibitem{Kingston07}
D.~B. Kingston, ``Decentralized control of multiple {UAV}s for perimeter and
  target suiveillance,'' \emph{Ph.D. dissertation, Brigham Young University},
  2007.

\bibitem{ShamesDFA12}
I.~Shames, S.~Dasgupta, B.~Fidan, and B.~D.~O. Anderson, ``Circumnavigation
  using distance measurements under slow drift,'' \emph{{IEEE} Transactions on
  Automatic Control}, vol.~57, no.~4, pp. 889--903, 2012.

\bibitem{DeghatSAY13}
M.~Deghat, I.~Shames, B.~D.~O. Anderson, and C.~Yu, ``Localization and
  circumnavigation of a slowly moving target using bearing measurements,''
  \emph{{IEEE} Transactions on Automatic Control}, 2013.

\bibitem{Warwick11}
\BIBentryALTinterwordspacing
G.~Warwick, ``Lightsquared tests confirm {GPS} jamming,'' \emph{Aviation Week},
  June 2011. [Online]. Available:
  \url{http://www.aviationweek.com/aw/generic/story.jsp?id=news/awx/2011/06/09/awx_06_09_2011_p0-334122.xml}
\BIBentrySTDinterwordspacing

\bibitem{ShepardBhattiHumphreys12}
\BIBentryALTinterwordspacing
D.~Shepard, J.~Bhatti, and T.~Humphreys, ``Drone hack: Spoofing attack
  demonstration on a civilian unmanned aerial vehicle,'' \emph{GPS World},
  2012. [Online]. Available: \url{http://www.gpsworld.com/drone-hack/}
\BIBentrySTDinterwordspacing

\bibitem{SahinogluGezici06}
Z.~Sahinoglu and S.~Gezici, ``Ranging in the {IEEE} 802.15.4a standard,'' in
  \emph{IEEE Annual Wireless and Microwave Technology Conference}, Clearwater
  Beach, FL, December 2006, pp. 1--5.

\bibitem{MatveevTeimooriSavkin09}
A.~S. Matveev, H.~Teimoori, and A.~V. Savkin, ``The problem of target following
  based on range-only measurements for car-like robots,'' in \emph{Joint IEEE
  Conference on Decision and Control and Chinese Control Conference}, Shanghai,
  China, December 2009, pp. 8537--8542.

\bibitem{ChaudharySinha12}
G.~Chaudhary and A.~Sinha, ``Detecting a target location using a mobile robots
  with range only measurements,'' in \emph{7th IEEE Conference on Industrial
  Electronics and Applications}, Singapore, July 2012, pp. 28--33.

\bibitem{SlotineLi91}
J.-J.~E. Slotine and W.~Li, \emph{Applied Nonlinear Control}.\hskip 1em plus
  0.5em minus 0.4em\relax Englewood Cliffs, NJ: Prentice Hall, 1991.

\end{thebibliography}

\end{document}